\documentclass[aip,
]{revtex4-1}

\usepackage{graphicx}
\usepackage{dcolumn}
\usepackage{bm}

\usepackage{siunitx}
\usepackage{color}
\usepackage{ulem}

\begin{document}

\title{Direct phase mapping of broadband Laguerre-Gaussian metasurfaces}

\author{Alexander Fa{\ss}bender}
\affiliation{Physikalisches Institut, Nussallee 12, Universit\"at Bonn, 53115 Bonn, Germany}

\author{Ji{\v{r}}{\'i} Babock{\'y}}
\affiliation{Central European Institute of Technology, Brno University of Technology, Purky\v{n}ova 123, 612 00 Brno, Czech Republic}

\author{Petr Dvo\v{r}\'{a}k}
\affiliation{Central European Institute of Technology, Brno University of Technology, Purky\v{n}ova 123, 612 00 Brno, Czech Republic}

\author{Vlastimil K{\v{r}}{\'a}pek}
\affiliation{Central European Institute of Technology, Brno University of Technology, Purky\v{n}ova 123, 612 00 Brno, Czech Republic}

\author{Stefan Linden}
\email{linden@physik.uni-bonn.de}
\affiliation
{Physikalisches Institut, Nussallee 12, Universit\"at Bonn, 53115 Bonn, Germany}

\date{\today}

\begin{abstract}
We report on the fabrication of metasurface phase plates consisting of gold nanoantenna arrays that generate Laguerre-Gaussian modes from a circularly polarized Gaussian input beam. 
The corresponding helical phase profiles with radial discontinuities are encoded in the metasurfaces by the orientation of the nanoantennas. 
A common-path interferometer is used to determine the orbital angular momentum of the generated beams. Additionally, we employ digital holography to record detailed phase profiles of the Laguerre-Gaussian modes. Experiments with different laser sources demonstrate the broadband operation of the  metasurfaces.

\end{abstract}

                             
\maketitle

Optical vortex beams have been the subject of intense research activities in recent years\cite{Allen1992,Yao2011}
and have found numerous applications in optical micromanipulation\cite{He1995}, quantum optics\cite{Mair2001}, imaging\cite{He1995}, and communications\cite{Willner2015}. 
Their characteristic feature is a helical phase distribution with a phase singularity on the optical axis. 
As a consequence of this, optical vortex beams possess annular intensity cross sections with strictly zero on-axis intensity.
Moreover, they carry an orbital angular momentum (OAM) of $\hbar l$ per photon, which is independent of the polarization state of the beam.
Here, $l$ is the so-called topological charge that determines the number and handedness of the intertwined helical phase fronts of the vortex.
Prime examples of optical vortex beams are Laguerre-Gaussian modes ($LG_{l,p}$) with azimuthal index $\vert l\vert \ge 1$ \ and arbitrary radial index $p$ \cite{Allen1992}.

Optical vortex beams can be generated by a number of different methods, e.g., astigmatic mode converters\cite{Allen1992}, spiral phase plates\cite{Sueda2004}, spatial light modulators\cite{Ohtake2007,Matsumoto2008}, and diffraction gratings\cite{Bazhenov1992}.
A promising new approach is based on metasurfaces\cite{Yu2011,Genevet2012,Huang2012,Maguid2016,Huang2017,Wang2017}. A metasurface is an artificial ultrathin optical device that consists of a dense arrays of sub-wavelength building-blocks \cite{Yu2014}.
These so-called meta-atoms serve as light scattering elements with properties that can be engineered by their geometry and material composition.
A light beam impinging on the metasurface interacts with the meta-atoms and gives rise to a scattered wave. 
The resulting wave front is thereby determined by the spatial variation of the scattering properties of the metasurface. 
By encoding an azimuthal phase factor $\exp\left(\imath \varphi l\right)$ into the metasurface, one can generate an optical vortex beam with topological charge $l$ from a Gaussian input beam. 
In particular, geometric metasurfaces based on the Pancharatnam-Berry phase concept\cite{Huang2012,Maguid2016,Wang2017} are suited for this purpose, as they combine a simple design principle with broadband operation and ease of fabrication.
This type of metasurface employs simple dipole antennas, e.g., plasmonic nanorods, as scattering elements to partially convert a circularly polarized input beam into light with the opposite circular polarization. The phase shift $\phi$ introduced by one of the dipoles is given by $\phi=2 \sigma \theta$, where $\theta$ is the angle enclosed between the dipole axis and a reference axis (in our case the $x$-axis) and $\sigma$ characterizes the circular polarization state of the incident light (right circular polarization (RCP): $\sigma=1$, left circular polarization (LCP): $\sigma=-1$). Based on this simple rule, the desired phase distribution $\phi(x,y)$ can be directly translated into the required orientation $\theta(x,y)$ of the meta-atom at the position $(x,y)$ in the metasurface.
\begin{figure}[hbt]
	\centering
    \fbox{\includegraphics[width=0.75\linewidth]{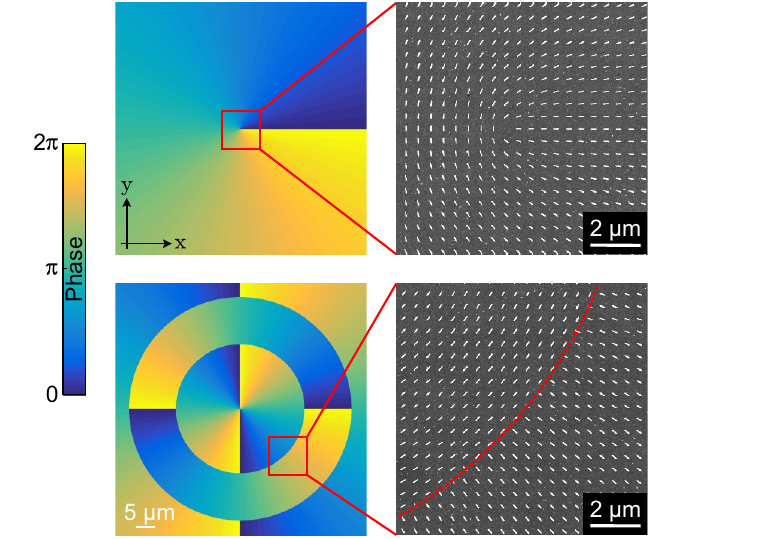}}
	\caption{The upper images show the phase distributions for a $LG_{1,0}$ mode with a sector of the metasurface captured by an electron micrograph. The lower images depict a $LG_{2,2}$ mode, designed for a beam waist radius of $w_0=\SI{17}{\micro\metre}$.}
	\label{Fig_1}
\end{figure}

In this article, we report on the generation of Laguerre-Gaussian beams using geometric metasurfaces consisting of gold nanorod antennas.
A circularly polarized Gaussian input beam is transmitted through a metasurface to imprint the phase distribution $\phi_{l,p}(r,\varphi)$ of the desired $LG_{l,p}$ beam onto the scattered wave with the opposite circular polarization:
\begin{equation}\label{eq1}
\phi_{l,p}(r,\varphi)=\varphi l + \pi u\left[-L^{\vert l \vert}_p\left(2 r^2/w_0^2\right)\right].
\end{equation}
Here, $u(x)$ is the unit step function, $L^{l}_{p}(x)$ is a generalized Laguerre polynomial, and $w_0$ is the waist radius of the incident Gaussian beam.
The first addend $\varphi l$ in equation (\ref{eq1}) is responsible for the helical phase profile, while the second addend accounts for the phase jumps of the $LG_{l,p}$ beam in radial direction.

\begin{figure}[h!]
	\centering
    \fbox{\includegraphics[width=0.75\linewidth]{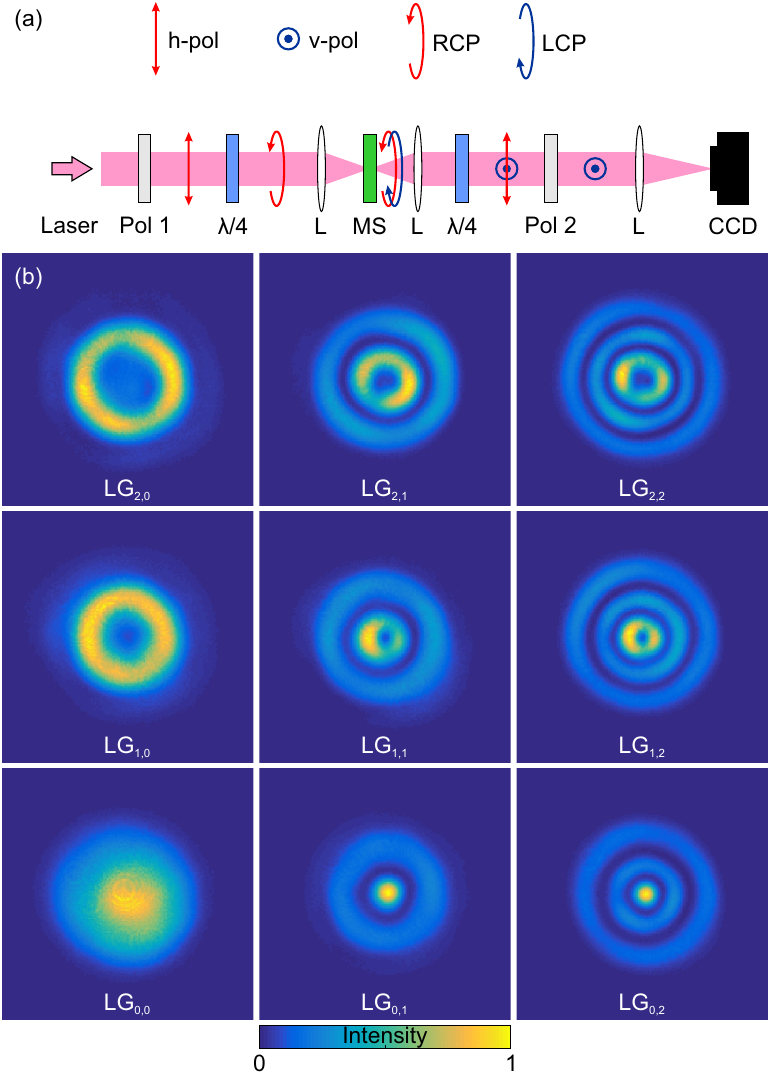}}
	\caption{(a) Scheme of the setup used for the intensity measurements. Polarizer 2 can be tuned azimuthally to create a common-path interferometer. (b) Measured intensity distributions of different Laguerre-Gaussian modes.}
	\label{Fig_2}
\end{figure}

The geometric metasurface phase plates are fabricated on top of an indium tin oxide covered glass substrate by standard electron beam lithography in combination with thermal evaporation of gold. 
Each metasurface has a circular shape with a diameter of $\SI{100}{\micro \metre}$ and consists of a square array of gold nanorods with a period of $\SI{500}{\nano \metre}$.
The dimensions of a nanorod are $\SI{220}{\nano \metre} \times \SI{60} {\nano \metre} \times \SI{40}{\nano \metre}$ (length $\times$ width $\times$ height). 
For these parameters, the nanorods support a localized plasmon mode with a resonance wavelength of $\SI{1080}{\nano\metre}$. 
The orientation of the nanorods in a given metasurface encodes the desired phase distribution. 
For the generation of a Laguerre-Gaussian beam with indices $l$ and $p$, the angle $\theta(r,\varphi)$ between the long axis of the nanorod at the position ($r$, $\varphi$) and the $x$-axis is chosen to be $\theta=\phi_{l,p}(r,\varphi)/2$.
Figure\,\ref{Fig_1} depicts scanning electron micrographs of sections of two metasurface phase plates  together with the corresponding phase distributions.


The intensity profiles of the generated Laguerre-Gaussian beams are characterized with the setup schematically shown in Fig.\,\ref{Fig_2}a. 
As light source we use a continuous wave diode laser with a wavelength of $\lambda=\SI{780}{\nano \metre}$ that is spatially filtered by a single mode fiber to guarantee a high quality Gaussian beam.  
The input beam is sent through a circular polarizer consisting of a linear polarizer and a quarter wave plate and focused ($f=\SI{100}{\milli \metre}$) onto the phase plate (waist radius $w_0=\SI{17}{\micro \metre}$).
The scattered light as well as the transmitted Gaussian beam are collected with a second lens ($f=\SI{100}{\milli \metre}$).
A crossed circular analyzer blocks the input beam and the transmitted Laguerre-Gaussian beam is imaged onto a CCD camera.
Figure\,\ref{Fig_2}b shows intensity distributions of several Laguerre-Gaussian beams. Intensity distributions of beams with the same azimuthal (radial) index are arranged in the same line (column).
In this figure, the radial mode index $p$ increases successively by one from left to right. As expected, the number of radial discontinuities in the corresponding intensity distributions (dark rings around the center) increases likewise. The azimuthal mode index $l$ grows in steps of one from bottom to top. The associated increase in topological charge becomes noticeable by the growing low intensity region centered around the beam axis. 

\begin{figure}[hbt]
	\centering
    \fbox{\includegraphics[width=0.75\linewidth]{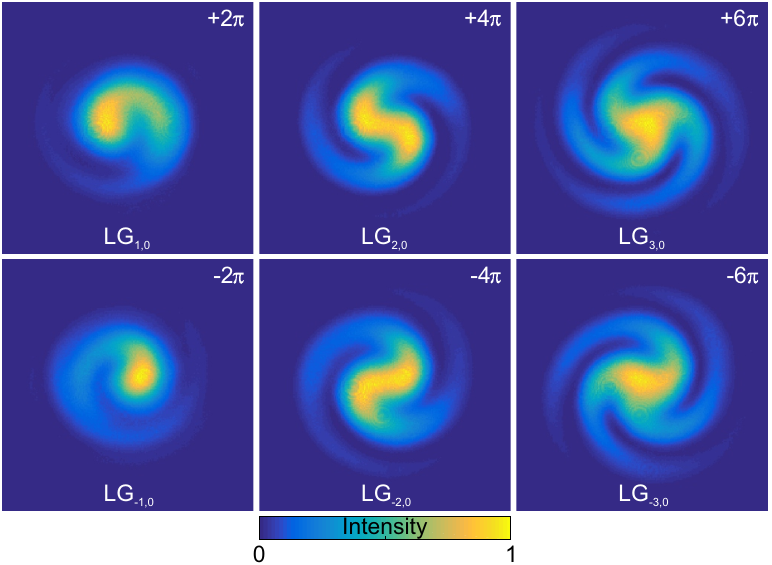}}
	\caption{Results of the interferometric measurements, performed with a common-path interferometer. The OAM carried by these six different metasurfaces can be easily identified due to the overlap with a Gaussian beam of $l=0$. The number in each upper right corner indicates the phase shift that is introduced in one azimuthal turn around the beam axis.}
	\label{Fig_3}
\end{figure}

To measure the topological charge of the generated Laguerre-Gaussian beams, we use the setup as a common-path interferometer.
The transmitted Gaussian beam and the generated Laguerre-Gaussian beam have orthogonal linear polarizations after passing the second quarter-wave plate (see Fig.\,\ref{Fig_2}a). By adjusting the polarization axis of polarizer 2, we can overlap both fields on the CCD with comparable amplitudes. The resulting interference image allows for an easy and fast determination of the topological charge of the Laguerre-Gaussian beam. Since the Gaussian input has a flat phase front ($l=0$), we can directly deduce the absolute value $\vert l\vert$ of the topological charge of the generated Laguerre-Gaussian beam from the number of spiral fringes in the interference image. The sign of the topological charge follows from the handedness of the spirals. Exemplary interference images of Laguerre-Gaussian beams with $p=0$ and $l=\pm1, \pm2, \pm3$ are depicted in Fig.\,\ref{Fig_3}. 

\begin{figure}[h!]
	\centering
    \fbox{\includegraphics[width=0.75\linewidth]{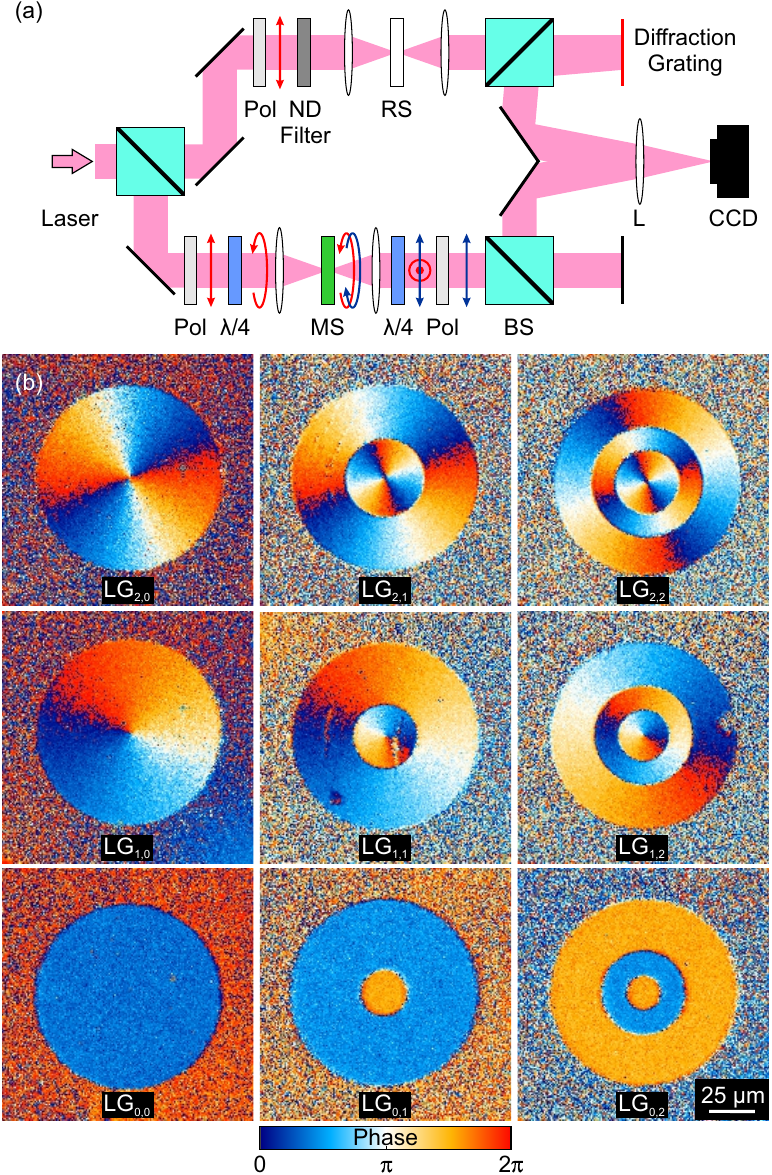}}
	\caption{(a) Scheme of the CCHM, consisting of reference arm with reference sample (RS) and sample arm with metasurface (MS). (b) Detailed phase profiles of the Laguerre-Gaussian modes ($0\leq l\leq2$ and $0\leq p\leq2$) recorded with a holographic microscope.}
	\label{Fig_4}
\end{figure}
To record detailed phase profiles of the generated Laguerre-Gaussian beams directly behind the metasurface phase plates, we employ a coherence-controlled holographic microscope (CCHM). This microscope allows the use of spatial and temporal incoherent light source for illumination \cite{Slaby2013,Babocky2017}. While typical off-axis systems cannot operate with incoherent light, the CCHM can handle the introduced incoherence. Therefore, it holds the advantage of an in-line system utilizing incoherent light for the suppression of coherent noise and combines it with the advantage of an off-axis system, which needs only one single interferographic image to extract the amplitude and phase distribution in detail.\\
A halogen lamp with a bandpass filter (center wavelength $\SI{650}{\nano \metre}$) can thus be used as light source. The beam is split up into reference and sample arm, see Fig.\,\ref{Fig_4}a. Polarization optics let only pass right handed circularly polarized light. Both arms are equipped with the same objectives for focusing onto the substrate and recollecting the light. The metasurface lies in the focal plane between the microscope objectives of the sample arm, while the reference arm contains a glass plate to adjust for the optical path through the substrate. Additionally, polarization optics in the sample arm behind the phase plate filter out the transmitted incident beam. A diffraction grating in the reference arm can correct for the incoherent illumination and is imaged onto the output plane. The hologram resulting from the overlap of sample and reference arm is recorded by a camera. The beams enclose an angle. Amplitude and phase of the beam behind the metasurface can be reconstructed by the method of digital holography\cite{Cuche1999}. 
Due to imperfect alignment and depolarization effects, the light transmitted through the sample arm still contains a contribution of the driving field. To retrieve solely the field scattered by the metasurface, we measured and reconstructed the field transmitted through a non-active part of the sample (i.e., without nanorods), which represents the portion of the driving field transmitted due to imperfect setup. This field was subsequently numerically subtracted from the total field transmitted through the active part of the sample. Figure\,\ref{Fig_4}b shows the phase reconstructed from CCHM measurements for metasurfaces generating various Laguerre-Gaussian modes. Clearly, there is very good agreement with designed phase distribution. \\

\begin{figure}[hbt]
	\centering
    \fbox{\includegraphics[width=0.75\linewidth]{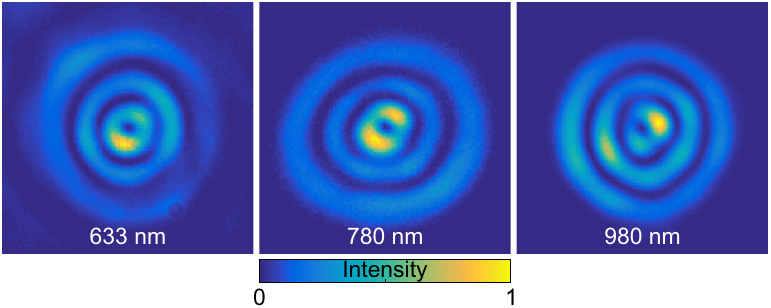}}
	\caption{Broadband application of a $LG_{1,2}$ metasurface, demonstrated at wavelengths of $\SI{633}{\nano \metre}$, $\SI{780}{\nano \metre}$, and $\SI{980}{\nano \metre}$, respectively.}
	\label{Fig_5}
\end{figure}

The experiment can also be conducted at different wavelengths. We additionally chose wavelengths of $\lambda_2=\SI{633}{\nano \metre}$ and $\lambda_3=\SI{980}{\nano \metre}$ to illustrate the broadband working regime of the metasurface. None of these sources drives the rods in resonance. Figure\,\ref{Fig_5} depicts the same metasurface arrangement ($LG_{1,2}$) illuminated by the three different light sources. The measurements show no qualitative differences. 

In conclusion, we fabricate metasurfaces that imprint the phase profiles of Laguerre-Gaussian modes on a Gaussian beam. The nanorods forming the metasurface are plasmonic antennas that introduce the necessary phase shift by the scattering of light into the opposite circular polarization. 
Based on the principle of the geometric Pancharatnam-Berry phase, the phase profile is encoded in the orientation of the nanorods. The off-resonant scattering of the nanoantennas allows for a broadband application. We demonstrate the operation at three different wavelength: $\SI{633}{\nano \metre}$, $\SI{780}{\nano \metre}$, and $\SI{980}{\nano \metre}$. The value and sign of the quantum number of orbital angular momentum are measured using a common-path interferometer. Detailed phase profiles are obtained from a single holographic image using coherence-controlled holographic microscopy.

The authors declare no competing financial interest.
S.L. acknowledges financial support by the German Federal Ministry of Education and Research through the funding program Photonics Research Germany (project 13N14150). CCHM measurements were carried out with support of the Ministry of Education, Youth and Sports of the Czech Republic (projects CEITEC 2020, No.~LQ1601, and CEITEC Nano RI, No.~LM2015041).

\bibliography{thesis_refs} 

\end{document}